\documentclass[smallextended]{svjour3}
\usepackage{graphicx}
\usepackage{epstopdf}
\graphicspath{ {images/} }
\usepackage{amsmath}
\usepackage{amsfonts} %type mathematics
\usepackage{amssymb} %type mathematics, same above
\usepackage{bm}
\usepackage{float}
\usepackage[numbers]{natbib}
\usepackage{color}
\usepackage{background}
\SetBgContents{The final publication is available at Springer http://dx.doi.org/10.1007/s11277-016-3865-9  (Read Online at http://rdcu.be/qmtg)}
\SetBgScale{1}
\SetBgAngle{0}
\SetBgOpacity{1}
\SetBgPosition{current page.south}
\SetBgVshift{2cm}

\begin{document}

\title{Enhancement of Physical Layer Security \\ Using Destination Artificial Noise \\ Based on Outage Probability}

\author{Ali Rahmanpour \and Vahid T. Vakili \and S. Mohammad Razavizadeh}
\institute{The authors are with the School of Electrical Engineering, Iran University of Science and Technology (IUST), Narmak, Tehran 1684613114, Iran.
\email{rahmanpour@alumni.iust.ac.ir \and vakily@iust.ac.ir \and smrazavi@iust.ac.ir.}}


\maketitle

\begin{abstract}
%\boldmath
In this paper, we study using Destination Artificial Noise (DAN) besides Source Artificial Noise (SAN) to enhance physical layer secrecy with an outage probability based approach. It is assumed that all nodes in the network (i.e. source, destination and eavesdropper) are equipped with multiple antennas. In addition, the eavesdropper is passive and its channel state and location are unknown at the source and destination. In our proposed scheme, by optimized allocation of power to the SAN, DAN and data signal, a minimum value for the outage probability is guaranteed at the eavesdropper, and at the same time a certain level of signal to noise ratio (SNR) at the destination is ensured. Our simulation results show that using DAN along with SAN brings a significant enhancement in power consumption compared to methods that merely adopt SAN to achieve the same outage probability at the eavesdropper.
\end{abstract}


\section{Introduction}
Due to weakness of traditional security methods which are based on using cryptography algorithms in upper layers, physical layer security attracts many attentions in the recent years.
Physical Layer Security was first addressed by Wyner's celebrated paper \cite{Wiretap:1975} in which a wiretap channel was studied and the notion of secrecy capacity was introduced. Since then, several studies have extended the Wyner's work, such as \cite{Gaussian:1978} for Gaussian wiretap channels, \cite{SecureFading:2008} for fading channels and \cite{SecrecyMIMO:2011} for Multiple-Input Multiple-Output (MIMO) channels.

One practical method in this area is using beamforming techniques in combination with emission of an Artificial Noise (AN) at the data source side (Alice) to corrupt an eavesdropper's reception during transmission of data to the destination (Bob) \cite{Guaranteeing:2008}. During past years several works have developed this idea that most of them aim to maximize secrecy capacity. In a different way, the authors in \cite{Outage:2012} proposed a probability based approach for employing AN to guarantee a minimum level of outage probability at an unknown Eve while satisfying a certain Quality of service (QoS) requirement at the Bob. It should be noted that all these works are based on exploiting Source Artificial Noise (SAN) that means sending a noise-like signal at the data-source side of a wireless link.

Recently, thanks to the development of full-duplex communications, one node can transmit and receive data signals at the same time and the same frequency band \cite{Co-Channel:2010}, \cite{Breaking:2013}. By full-duplex communications, it is possible to adopt Destination Artificial Noise (DAN) at the destination along with the SAN at the source. This idea was first proposed in \cite{SecureCommunication:2012} and then in \cite{Improving:2013} for Single-Input Multiple-Output (SIMO) communication systems. In \cite{Securing:2013} the MIMO communication systems when only one of the receiver antennas is used for data reception has been studied, but in \cite{Application:2014} destination can allocate more antennas for receiving the data. Similar to SAN, most works on the DAN are also based on maximizing secrecy capacity.

In this paper unlike to the previous works, we adopt an outage probability based approach for the characterization of the physical layer security that uses both DAN and SAN. In addition, we propose an optimal power allocation method to the DAN, SAN and source information signal to ensure the security requirements. We also investigate the effect of considering a power constraint at the destination for the situation in which Bob can only cancel a certain amount of self-interference. It is shown how this power constraint increases the total required power for a certain level of security. 

Our simulation results show that the proposed method brings a significant reduction in the total power consumption compared to the previous works where only SAN is used to ensure a certain outage probability at the Eve. We also represent the effect of Eve's location on the performance of the proposed method. It is shown that in contrast to previous works, we are able to ensure secrecy requirements even where Eve is very close to the Bob. 

This paper is organized as follow. After presenting system model in Section 2, the power allocation problem is proposed in Section 3. In Section 4, simulation results are presented for both constrained and unconstrained power scenarios at the destination. Finally in Section 5, the paper has been concluded.

\textit{Notation}: Bold symbols in small and capital letter denote vectors and matrices. In addition $\boldsymbol{(.)}^{H}$ denotes the conjugate transpose and $\left \| \boldsymbol{.} \right \|$ is the norm operator.

\section{SYSTEM MODEL}
In this paper, we consider a network consisting of a data source (Alice), a destination (Bob) and an eavesdropper (Eve). It is assumed that the source and the destination are equipped with $N_{A}\geq2$ and $N_{B}\geq2$ antennas, respectively, while Eve is equipped with $N_{E}\geq1$ antenna. The channel gains between Alice and Bob is represented by a matrix $\boldsymbol{H}_{AB}$ of size $N_{A}\times N_{B}$ and it is assumed to be known to all nodes. In the other hand the channel gains between Alice and Eve is represented by a matrix $\boldsymbol{H}_{AE}$ of size $N_{A}\times N_{E}$ and the channel gains between Bob and Eve by a matrix $\boldsymbol{H}_{BE}$ of size $N_{B}\times N_{E}$ which are unknown for the legal nodes (Alice and Bob). All channel gains are modelled by independent zero-mean complex Gaussian random variables. The variances of the above channels are $\sigma_{H_{AB}}^{2}$, $\sigma_{H_{AE}}^{2}$ and $\sigma_{H_{BE}}^{2}$ for the channel between Alice and Bob, Alice and Eve and Bob and Eve, respectively. If $\boldsymbol{H}$ be a channel gains matrix (small scale fading effects), by considering the \emph{path loss} effect, we can define $\hat{\boldsymbol{H}}$ as \cite{Goldsmith},\cite{Application:2014}
\begin{equation} \label{eq:Hhat}
\hat{\boldsymbol{H}}= (\lambda_{0} r^{-\kappa})^{1/2} \boldsymbol{H}
\end{equation}
where $\lambda_{0}$ is a constant that is used for showing the power at a reference signal and it determined by empirical measurements. In addition $r$ is the distance between two nodes and $\kappa$ is the path loss exponent and depends on the propagation environment. The values of this parameter is typically between 2 and 6.

\begin{figure}[t]
%\centering 
\includegraphics[width=\columnwidth]{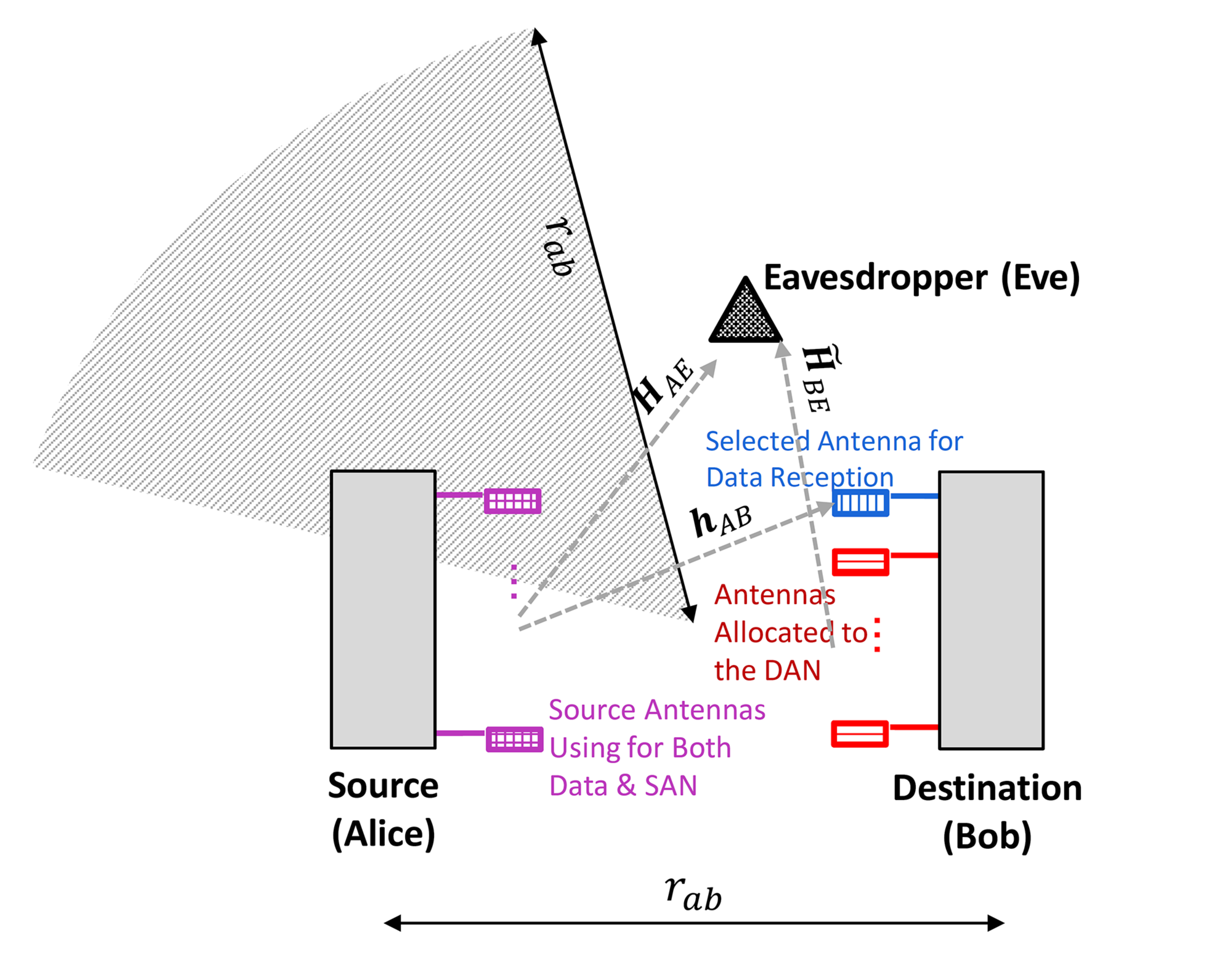}
\caption{System model}
\label{fig:model}
\end{figure}

We assume that the Eve's location is unknown to the legal nodes and has a uniform distribution in a circle with radius $r_{ab}$, where $r_{ab}$ is the distance between Alice and Bob. Also it is assumed that the legitimate users are located at the center of the area (Fig.~\ref{fig:model}).

%\ref{fig:mesh1}

To enhance the security of the system, we assume that only one of the destination antennas is employed for receiving the data signal and the other antennas are assigned to the AN propagation (i.e. DAN). On the other hand, all source antennas could be used to transmit both precoded data signal and AN (i.e. SAN). To decrease the power consumption and enhance the performance at the Bob, we select its best antenna (i.e. the \textit{j}th antenna) for receiving data from Alice. One approach for selecting this antenna is as follows: 
\begin{equation} \label{i-arg}
j=\arg \max_{i} \left \{ \left \| \boldsymbol{H}^{i}_{AB} \right \| \right \}
\end{equation}
%\underset{i}
where $\boldsymbol{H}^{i}_{AB}$ is the \textit{i}th column of $\boldsymbol{H}_{AB}$ which is related to the \textit{i}th antenna at the destination and $i=1,2,...,N_{B}$. Therefore the channel assigned to the data transmission between Alice and Bob is denoted by a vector $\boldsymbol{h}_{AB}$ of size $N_{A}$. In addition, the channel matrix between the remaining Bob's antennas and Eve is presented by a matrix $\boldsymbol{\tilde{H}}_{BE}$ of size $(N_{B}-1)\times N_{E}$. 

In addition assuming that Eve uses the selection combining method, the channel vector between Eve's \textit{k}th antenna and Alice is denoted by $\boldsymbol{h}_{AE}$ of size $N_{A}$, while $\boldsymbol{h}_{BE}$ of size $(N_{B}-1)$ denotes the channel between Bob's remained antennas and Eve's \textit{k}th antenna.

Now let $\boldsymbol{x}$, $y_{B}$ and $\boldsymbol{y}_{E}$ be the precoded data signal transmitted by Alice, the received signal at Bob and the received signal at Eve, respectively. Therefore
\begin{equation} \label{y:received}
y_{B}=\hat{\boldsymbol{h}}^{H}_{AB}\boldsymbol{x}+n_{B}
\end{equation}
\begin{equation} \label{z:received}
\boldsymbol{y}_{E}=\hat{\boldsymbol{H}}^{H}_{AE} \boldsymbol{x} + \hat{\boldsymbol{\tilde{H}}}^{H}_{BE} \boldsymbol{\nu} + \boldsymbol{n}_{E}
\end{equation}
where $\boldsymbol{\nu}$ is the DAN emitting by Bob and we assume that its self-interference can be completely canceled at the destination. Assuming there is no knowledge about the Eve's Channel State Information (CSI), DAN power must be equally distributed among $(N_{B}-1)$ Bob's antennas that are assigned to DAN propagation. Hence we can write
\begin{equation} \label{DAN}
\boldsymbol{\nu} = \sqrt{\frac{P}{N_{B}-1}}\sum_{\substack{i=1 \\ i\neq j}}^{N_{B}}\eta \boldsymbol{\psi}_{i}
\end{equation}
where $P\geq 0$ is the power allocated to DAN, $\eta$ is a complex scalar with unit magnitude and uniform random phase, $\boldsymbol{\psi}_{i}$ is a vector of size $N_{B}$ whose \textit{i}th element is equal to one and all other elements are equal to zero. $n_{B}$ and $\boldsymbol{n}_{E}$ of size $N_{E}$ are the Gaussian noise terms with zero mean and variances $\sigma_{B}^{2}$ and $\sigma_{E}^{2} \boldsymbol{I}$, respectively. $\boldsymbol{x}$ is the beamformed signal transmitted by Alice and could be written as
\begin{equation} \label{x-signal}
\boldsymbol{x}=\sqrt{\varphi \acute{P}}d\boldsymbol{t}+\sqrt{(1-\varphi ) \acute{P}} \boldsymbol{\eta}
\end{equation}
where $0\leq \varphi \leq 1$ is a parameter that determines the ratio of power assigned to the information signal and SAN, $\acute{P}>0$ is Alice's total power, $d$ is the information symbol with $\mathbb{E} \left \{ \left | d \right |^{2} \right \}=1$ and $\boldsymbol{t}$ of size $(N_{A})$ is normalized $(\left \| \boldsymbol{t} \right \|=1)$ beamforming vector. $\boldsymbol{\eta}$ is SAN vector of size $(N_{A})$ that is orthogonal to $\boldsymbol{t}$ $(i.e., \boldsymbol{t}^{H} \boldsymbol{\eta} =0)$ and its covariance  matrix is denoted by $\boldsymbol{C}_{\eta}$ as
\begin{equation} \label{covariance:SAN}
\boldsymbol{C}_{\eta}=\mathbb{E}\left \{ \boldsymbol{\eta} \boldsymbol{\eta}^{H} \right \}
\end{equation}
Therefore we have $Tr\left \{ \boldsymbol{C}_{\eta} \right \}=1$. For the beamforming at the source node, as proposed in \cite{Physical-Layer:2011}, if $\boldsymbol{t}_{i}$ is the \textit{i}th eigenvector of $\boldsymbol{h}_{AB} \boldsymbol{h}_{AB}^{H}$ and $\boldsymbol{t}_{1}$ is assumed to be the principal eigenvector, it is assumed that $\boldsymbol{t}=\boldsymbol{t}_{1}$. Based on the orthogonality of the eigenvectors of $\boldsymbol{h}_{AB} \boldsymbol{h}_{AB}^{H}$, $\boldsymbol{\eta}$ is a linear combination of $N_{A}-1$ eigenvectors, and hence it lies in the nullspace of  $\boldsymbol{h}_{AB}$. Since Eve's CSI is not known to the legal nodes, the noise power equally is distributed to these eigenvectors as follows
\begin{equation} \label{SAN}
\boldsymbol{\eta} =\sqrt{\frac{1}{N_{A}-1}}\sum _{i=2}^{N_{A}}\eta \boldsymbol{t}_{i}.
\end{equation}  

According to the above beamforming model, the signal to noise ratios (SNRs) at the Bob and Eve's \textit{k}th antenna can be derived as
\begin{equation} \label{SNR:B}
\mathrm{SNR_{B}}=\frac{\varphi \acute{P} \Vert \hat{\boldsymbol{h}}_{AB} \Vert^{2}}{\sigma_{B}^{2}}
\end{equation}
and
\begin{equation} \label{SNR:E}
\mathrm{SNR_{E}^{k}}=\frac{\varphi \acute{P}  \hat{\boldsymbol{h}}_{AE}^{H} \boldsymbol{t}_{1} \boldsymbol{t}_{1}^{H} \hat{\boldsymbol{h}}_{AE}  }{(1-\varphi ) \acute{P} \hat{\boldsymbol{h}}_{AE}^{H} \boldsymbol{C}_{\eta }  \hat{\boldsymbol{h}}_{AE}+\frac{P}{N_{B}-1} \hat{\boldsymbol{h}}_{BE}^{H} \hat{\boldsymbol{h}}_{BE}}.
\end{equation}

\section{The Power Allocation Problem}
In this section, we discuss about the problem of power allocation to SAN and DAN. To ensure secrecy, $\gamma _{b}$ and $\gamma _{e}$ QoS constraints should be satisfied at Bob and Eve and hence the power optimization problem can be written as:
\begin{subequations}
\label{eq:opt1}
\begin{align}
\min_{P, \acute{P}, \varphi } \quad & P+\acute{P} \\
\label{eq:opt1-2}
\mathrm{s.t.} \quad & \mathrm{SNR_{B}} \ge \gamma _{b} \\
\label{eq:opt1-3}
& \mathbb{P} \lbrack \mathrm{SNR_{E}} \le \gamma _{e}\rbrack \ge \beta.
\end{align}
\end{subequations}

In above optimization problem, (\ref{eq:opt1-2}) is for satisfying a certain level of SNR at Bob and (\ref{eq:opt1-3}) guarantees a minimum value of outage probability at Eve. 

As it can be seen, our object is a joint power optimization for source and destination. Considering that Eve uses the selection combining method and the channel matrix coefficients are independent, we have
\begin{equation} \label{P:SNR_Ek}
\mathbb{P} \lbrack \mathrm{SNR_{E}} \le \gamma _{e}\rbrack = \prod_{k=1}^{N_{E}} \mathbb{P} \lbrack \mathrm{SNR_{E}^{k}} \le \gamma _{e}\rbrack \\
= \mathbb{P} \lbrack \mathrm{SNR_{E}^{k}} \le \gamma _{e}\rbrack^{N_{E}}
\end{equation}
By substituting (\ref{SNR:B}), (\ref{SNR:E}) and (\ref{P:SNR_Ek}) into (\ref{eq:opt1}) and considering (\ref{eq:Hhat}), we have:
\begin{subequations} 
\label{eq:opt2}
\begin{align}
\label{eq:opt2-1}
& \min _{P, \acute{P}, \varphi } \quad  P+ \acute{P} \\
\label{eq:opt2-2}
& \mathrm{s.t.} \nonumber \\
& \quad  \varphi \acute{P} \ge \gamma _{b}\sigma _{B}^{2} (\lambda_{0} r_{AB}^{-\kappa})^{-1} \Vert {\boldsymbol{h}}_{AB}\Vert^{-2} \\
\label{eq:opt2-3}
& \mathbb{P} \lbrack \frac{\varphi \acute{P} (\lambda_{0} r_{AE}^{-\kappa})  {\boldsymbol{h}}_{AE}^{H} \boldsymbol{t}_{1} \boldsymbol{t}_{1}^{H} {\boldsymbol{h}}_{AE} }{(1-\varphi ) \acute{P} (\lambda_{0} r_{AE}^{-\kappa}) {\boldsymbol{h}}_{AE}^{H} \boldsymbol{C}_{\eta } {\boldsymbol{h}}_{AE}+\frac {P} {N_{B}-1} (\lambda_{0} r_{BE}^{-\kappa}) {\boldsymbol{h}}_{BE}^{H} {\boldsymbol{h}}_{BE}} \nonumber \\
& \le \gamma _{e} \rbrack  \ge \beta^{\frac{1}{N_{E}}}
\end{align}
\end{subequations}
We can rewrite (\ref{eq:opt2-3}) as
\begin{equation} \label{eq:opt3-3}
\mathbb{P} \lbrack \boldsymbol{h}^{H}_{AE} \boldsymbol{a} \boldsymbol{h}_{AE} + b \boldsymbol{h}_{BE}^{H} \boldsymbol{h}_{BE} \le \sigma _{E}^{2} \rbrack \ge \beta^{\frac{1}{N_{E}}}
\end{equation}
where $\boldsymbol{a}$ and $b$ are defined as below
\begin{subequations} 
\label{eq:AandB}
\begin{align}
\label{eq:a}
& \boldsymbol{a}= \acute{P} \lambda_{0} \bar{r}_{ae}^{-\kappa} (\varphi \gamma 
_{e}^{-1} \boldsymbol{t}_{1} \boldsymbol{t}_{1}^{H} - (1-\varphi ) \boldsymbol{C}_{\eta }) \\
\label{eq:b}
& b= - \frac {P} {N_{B}-1} \lambda_{0} \bar{r}_{be}^{-\kappa}  
\end{align}
\end{subequations}
where $\bar{r}_{ae}=\mathbb{E}[r_{ae}]$ and $\bar{r}_{be}=\mathbb{E}[r_{be}]$. The left side in (\ref{eq:opt3-3}) can be interpreted as a Cumulative Distribution Function (CDF). By defining 
\begin{equation} \label{Xhah}
X=\boldsymbol{h}^{H}_{AE} \boldsymbol{a} \boldsymbol{h}_{AE}
\end{equation}
\begin{equation} \label{Yhh}
Y=\boldsymbol{h}_{BE}^{H} \boldsymbol{h}_{BE}
\end{equation}
and $Z=X+bY$, we have
%\begin{subequations} 
\label{eq:Fz}
\begin{align}
%\label{eq:Fz-1}
F_{Z}(z)&= \mathbb{P} \lbrack X+bY \le z \rbrack \nonumber \\
& = \mathbb{P} \lbrack \boldsymbol{h}^{H}_{AE} \boldsymbol{a} \boldsymbol{h}_{AE} + b \boldsymbol{h}_{BE}^{H} \boldsymbol{h}_{BE} \le z \rbrack \nonumber \\
\label{eq:Fz-3}
& =\int_{-\infty}^{+\infty}{f_{Y}(y)}dy \int_{-\infty}^{z-by} {f_{X}(x)}dx 
\end{align}
%\end{subequations}
where $\int_{-\infty}^{z-by} {f_{X}(\tau)}d\tau=F_{X}(z-by)$ is the CDF of $X$. On the other hand $X$ is an indefinite Hermitian quadratic form for $x \ge 0$ \cite{distribution:2009,Outage:2012}, so $F_{X}(x)$ can be derived as \cite{distribution:2009}:
\begin{equation} \label{eq:Fx}
F_{X}(x)=u(x)+\frac{\alpha _{1}}{\vert \lambda _{1}\vert }e^{(\frac{-x}{\lambda _{1}})} u(\frac{x}{\lambda _{1}})
\end{equation}
where
\begin{equation} \label{eq:alpha}
\alpha_{1}=-\frac{\lambda _{1}}{(1-\frac{\lambda _{2}}{\lambda_{1}})^{(N_{A}-1)}}
\end{equation}
and
\begin{subequations} 
\label{eq:lambda}
\begin{align}
\label{eq:lambda-1}
& \lambda _{1}=\varphi \acute{P}\lambda _{0} \bar{r}_{ae}^{-\kappa }\gamma_{e}^{-1}\sigma _{H_{AE}}^{2}\ge 0 \\
\label{eq:lambda-2}
& \lambda _{2}=-(1-\varphi )\acute{P}\lambda _{0} \bar{r}_{ae}^{-\kappa }\sigma_{H_{AE}}^{2}(N_{A}-1)^{-1}\le 0.
\end{align}
\end{subequations}

In addition, $Y$ is sum of the squares of $2(N_{B}-1)$ independent normal random variables and has chi-squared ($\chi^{2}$) distribution as 
\begin{equation} \label{eq:fY}
f_{Y}(y;2(N_{B}-1))=\frac{(\frac{y}{\sigma_{H_{BE}}^{2}})^{(N_{B}-2)}.e^{(\frac{-y}{2\sigma_{H_{BE}}^{2}})}}
{\sigma _{H_{BE}}^{2}2^{(N_{B}-1)}\Gamma (N_{B}-1)} u(y).
\end{equation}

Therefore (\ref{eq:Fz-3}) can be rewritten as 
\begin{equation} \label{eq:Fz2}
F_{Z}(z)=\int_{0}^{+\infty }{f_{Y}(y) dy}\int_{-\infty}^{z-by}{f_{X}(x)}dx,
\end{equation}
Considering $(z-by) \geq 0$, substituting (\ref{eq:Fx}) and (\ref{eq:fY}) into (\ref{eq:Fz2}) and considering $\sigma_{H_{BE}}^{2}=1$, we have
%\begin{subequations} 
\label{eq:Fz3}
\begin{align}
% \label{eq:Fz3-1} 
F_{Z}(z)& =\int_{0}^{+\infty}{\frac{1}{2^{(N_{B}-1)}(N_{B}-2)!}y^{(N_{B}-2)}e^{-0.5y}dy} \nonumber \\ 
\label{eq:Fz3-2}
& +\int_{0}^{+\infty}{\frac{\alpha _{1}e^{(\frac{-z}{\lambda_{1}})}}{2^{(N_{B}-1)}(N_{B}-2)!\vert \lambda _{1}\vert}y^{(N_{B}-2)}e^{(\frac{b}{\lambda _{1}}-0.5)y}dy}
\end{align}
%\end{subequations}
After some manipulations, $F_{Z}(z)$ in (\ref{eq:Fz3-2}) derived as
\begin{equation} \label{eq:Fz4}
F_{Z}(z)=(1+\frac{(-1)^{N_{B}}e^{(\frac{-z}{\lambda_{1}})}}{2^{(N_{B}-1)}(1-\frac{\lambda _{2}}{\lambda_{1}})^{N_{A}-1}(\frac{b}{\lambda _{1}}-0.5)^{(N_{B}-1)}}).
\end{equation}

Substituting (\ref{eq:lambda}) and (\ref{eq:Fz4}) into (\ref{eq:opt2}), the optimization problem in (\ref{eq:opt2}) is simplified as
\begin{subequations} 
\label{eq:opt4}
\begin{align}
\label{eq:opt4-1}
& \min _{P, \acute{P}, \varphi } \quad P+ \acute{P} \\
\label{eq:opt4-2}
& \mathrm{s.t.} \quad \varphi \acute{P} \ge \gamma _{b}\sigma _{B}^{2} (\lambda_{0} r_{AB}^{-\kappa})^{-1} \Vert {\boldsymbol{h}}_{AB}\Vert^{-2} \\
\label{eq:opt4-3}
& (1-\frac{2^{(1-N_{B})}e^{(\frac{-\gamma _{e}\sigma _{E}^{2}}{\varphi \acute{P}\lambda 
_{0} \bar{r}_{ae}^{-\kappa }\sigma_{H_{AE}}^{2}})}}{(1+\frac{(1-\varphi )\gamma _{e}}{\varphi (N_{A}-1)})^{(N_{A}-1)}(\frac{P \bar{r}_{be}^{-\kappa }\gamma _{e}}{(N_{B}-1) \varphi \acute{P} \bar{r}_{ae}^{-\kappa }\sigma _{H_{AE}}^{2}}+0.5)^{(N_{B}-1)}}) \nonumber \ge \beta^{\frac{1}{N_{E}}}. \\
\end{align}
\end{subequations}

According to (\ref{eq:opt4-3}), it is obvious that the optimum value for (\ref{eq:opt4-2}) is  $\varphi \acute{P} = \gamma _{b}\sigma _{B}^{2} (\lambda_{0} r_{AB}^{-\kappa})^{-1} \Vert {\boldsymbol{h}}_{AB}\Vert^{-2}$, therefore the optimization method needs to only be applied to (\ref{eq:opt4-3}). We use standard numerical methods to solve the problem.

\section{Simulation and Numerical Results}
In this section, we evaluate the performance of our proposed method by computer simulations. In our simulations, it is assumed that $\sigma_{H_{AB}}^{2}=\sigma_{H_{AE}}^{2}=\sigma_{H_{BE}}^{2}=1$, $r_{AB}=2$ \emph{Km}, $\gamma_{E}=0.5$ and $\sigma^{2}_{B}=\sigma^{2}_{N}=4 \times 10^{-14}$. For the path loss, it is assumed that $\lambda_{0}=10^{-1}$ and $\kappa=3$. Also the channel between Alice and Bob, $\boldsymbol{H}_{AB}$ is generated randomly with a distribution of $\boldsymbol{H}_{AB} \sim \mathcal{CN}(0,\sigma_{H_{AB}}^{2} \boldsymbol{I})$.

To represent how changes in the main parameters of our problem influence the security and power consumption, we first consider a scenario in which the destination has no power constraint for DAN. Then we also investigate the effect of a power constraint on DAN at the destination.

\subsection*{Case a) Destination has no power constraint for DAN}
\begin{figure}[h]
%\centering 
\includegraphics[width=\columnwidth]{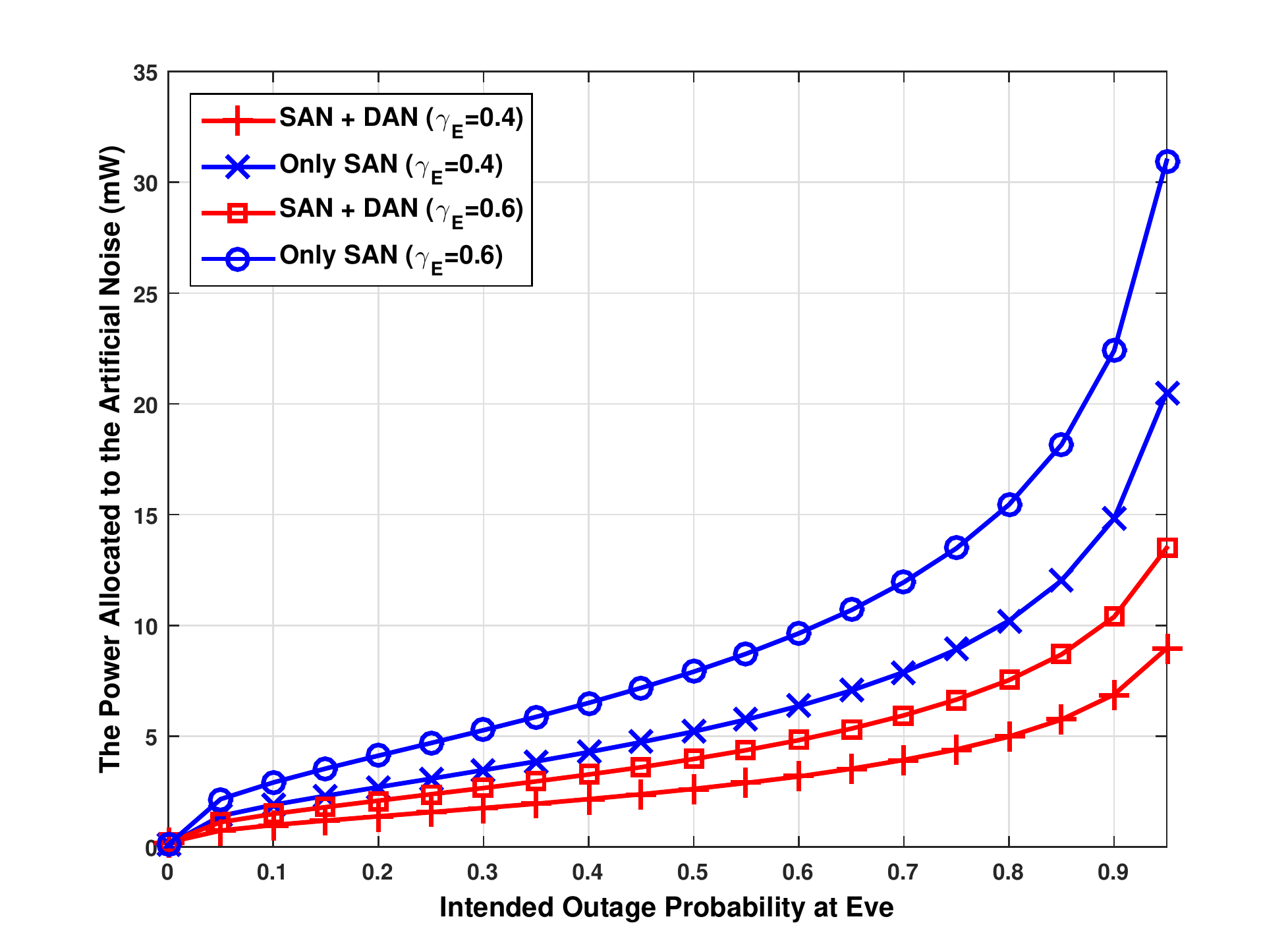}
\caption{The required power for AN (total SAN and DAN) versus the target outage probability ($\beta$) at the Eve, for $N_{E}=4$, for $\gamma_{E}=0.4$ and $0.6$.}
\label{fig:beta}
\end{figure}

As mentioned before, we assume that the Eve's location is distributed uniformly on the area and $\bar{r}_{ae}=\bar{r}_{be}=1000$. In addition we first assume that $N_{E}=N_{A}=N_{B}=4$. As it is seen in Fig.~\ref{fig:beta}, in the case that both DAN and SAN are used, the required power for AN to achieve a given outage probability at the Eve is much less than the case that only SAN is used. In addition, in both cases, with increasing $\gamma_{B}$, more power is needed to guarantee the given outage probability at the Eve. 

\begin{figure}[h]
%\centering 
\includegraphics[width=\columnwidth]{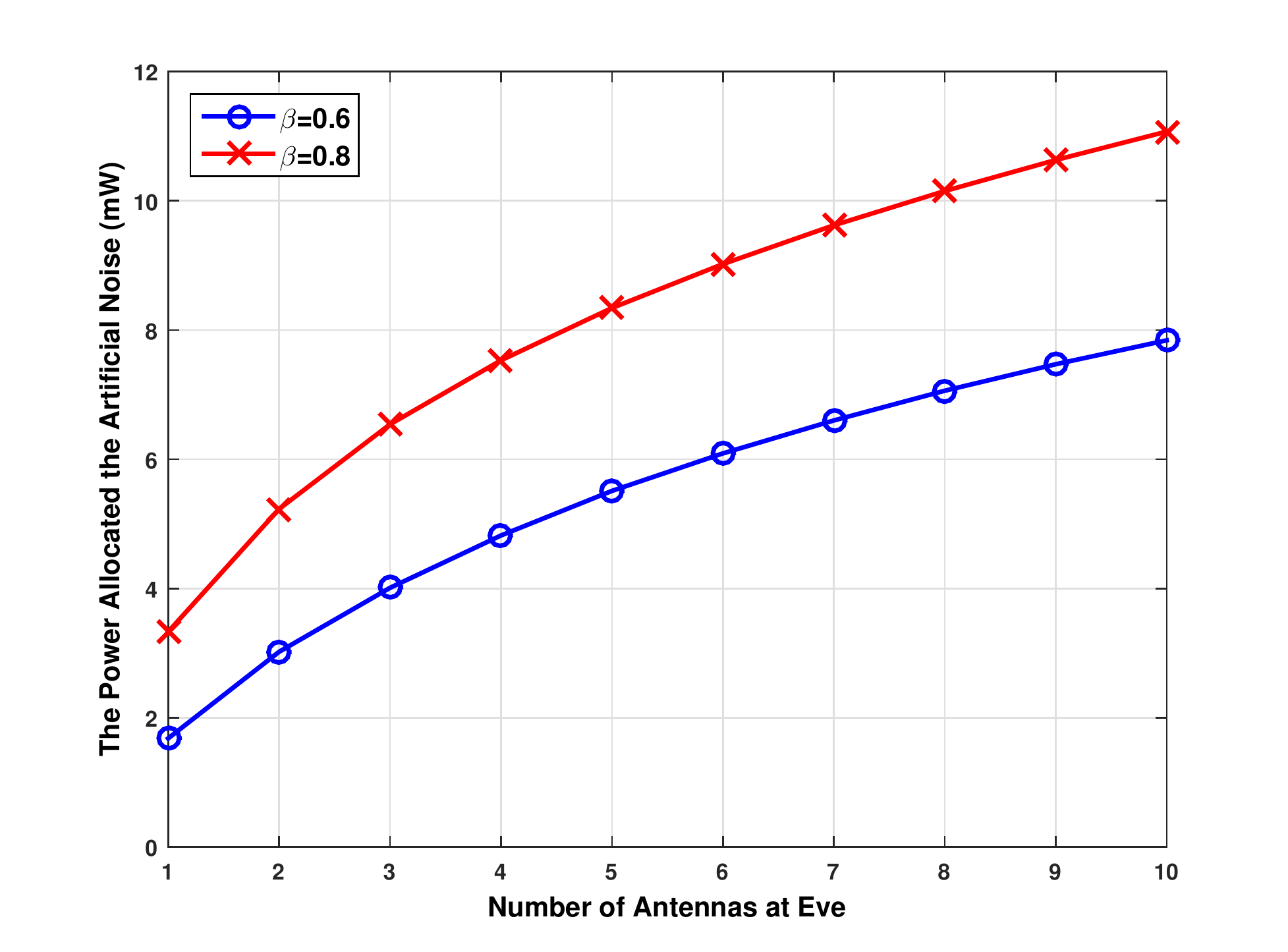}
\caption{The required power for AN versus the number of antennas at Eve, for $\beta=0.6$ and $0.8$.}
\label{fig:Ne}
\end{figure}

\begin{figure}[h]
%\centering 
\includegraphics[width=\columnwidth]{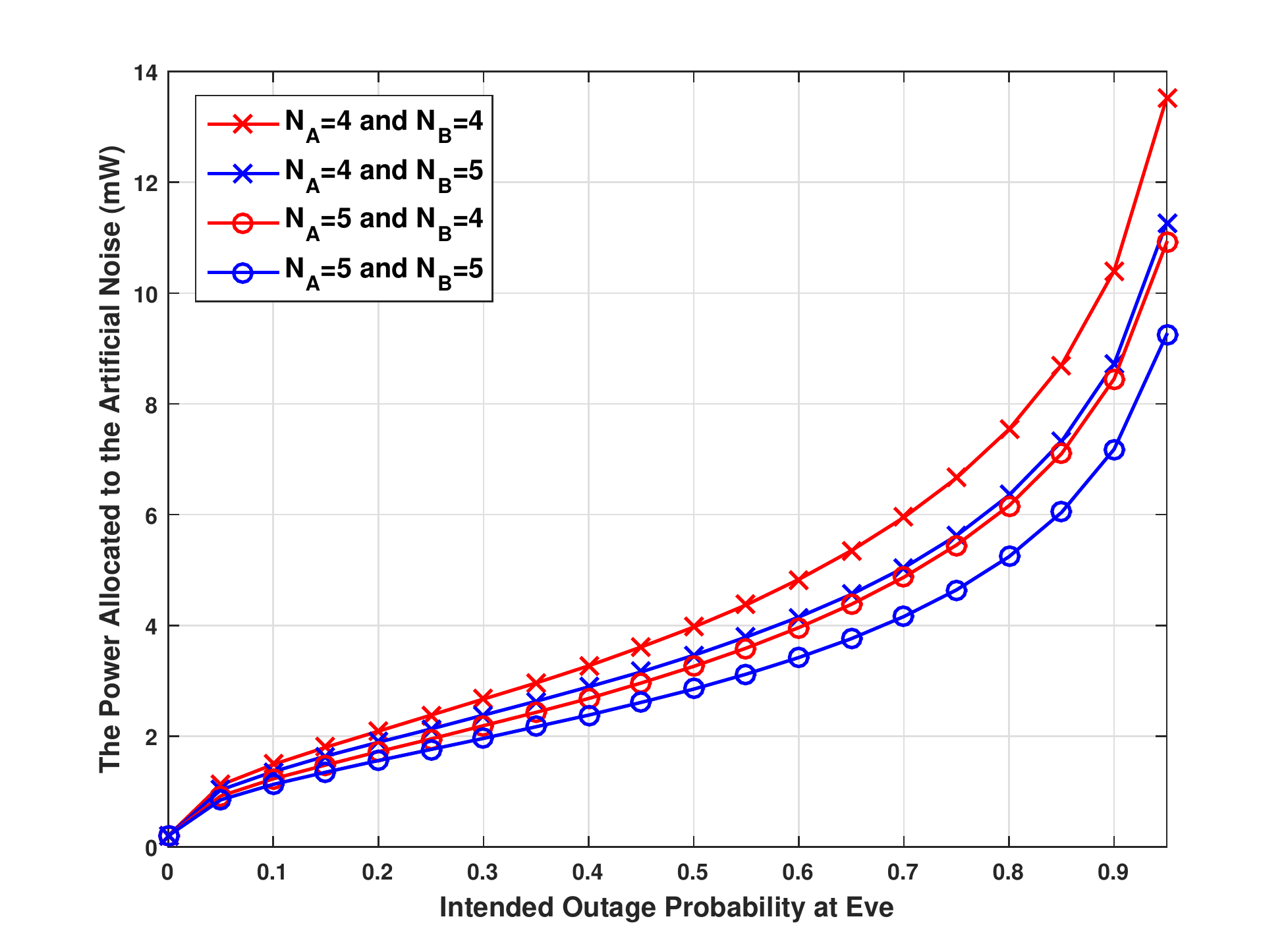}
\caption{The required power for AN versus the target outage probability ($\beta$) at the Eve, for different sets of $N_{A}$ and $N_{B}$ when $N_{E}=4$.}
\label{fig:Nab}
\end{figure}

In Fig.~\ref{fig:Ne}, we investigate the effect of different number of antennas at the Eve on the required power for the case $N_{A}=N_{B}=4$. It could be seen that when Eve has more capabilities, more power is needed to achieve a given outage probability. In Fig.~\ref{fig:Nab}, the required power for AN is shown for different values of $N_{A}$ and $N_{B}$, when $N_{E}=4$.

\begin{figure}[h]
%\centering 
\includegraphics[width=\columnwidth]{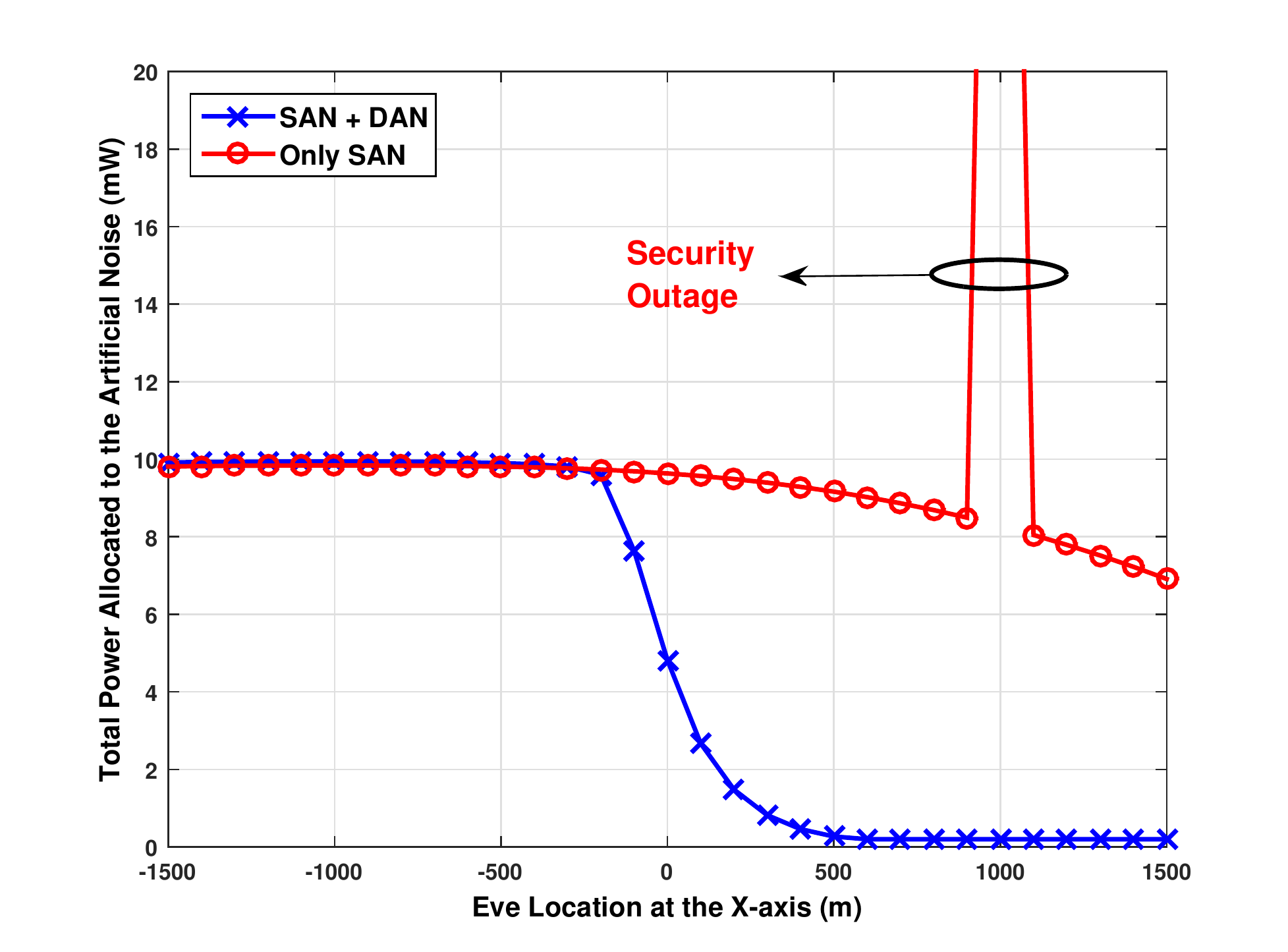}
\caption{The optimal power is needed for AN (total SAN and DAN) versus Eve's location changes from $(-1500,0)$ to $(1500,0)$, for $\beta=0.6$}
\label{fig:distance}
\end{figure}

In another scenario, we assume Alice and Bob are located at locations $(-1000,0)$ and $(1000,0)$, $N_{E}=N_{A}=N_{B}=4$, $\gamma_{B}=0.5$ and  $\beta=0.6$. The total power should be allocated to AN to achieve a target outage probability at Eve $(\beta)$ while Eve moves from $(-1500,0)$ to $(15000,0)$ is shown in Fig.~\ref{fig:distance}. From this figure, it is seen that by using DAN we need less power to guarantee a similar $\beta$. Especially in the case that Eve is close to Bob, it is still possible to guarantee a given security.

\subsection*{Case b) Destination has a power constraint}
\begin{figure}[h]
%\centering 
\includegraphics[width=\columnwidth]{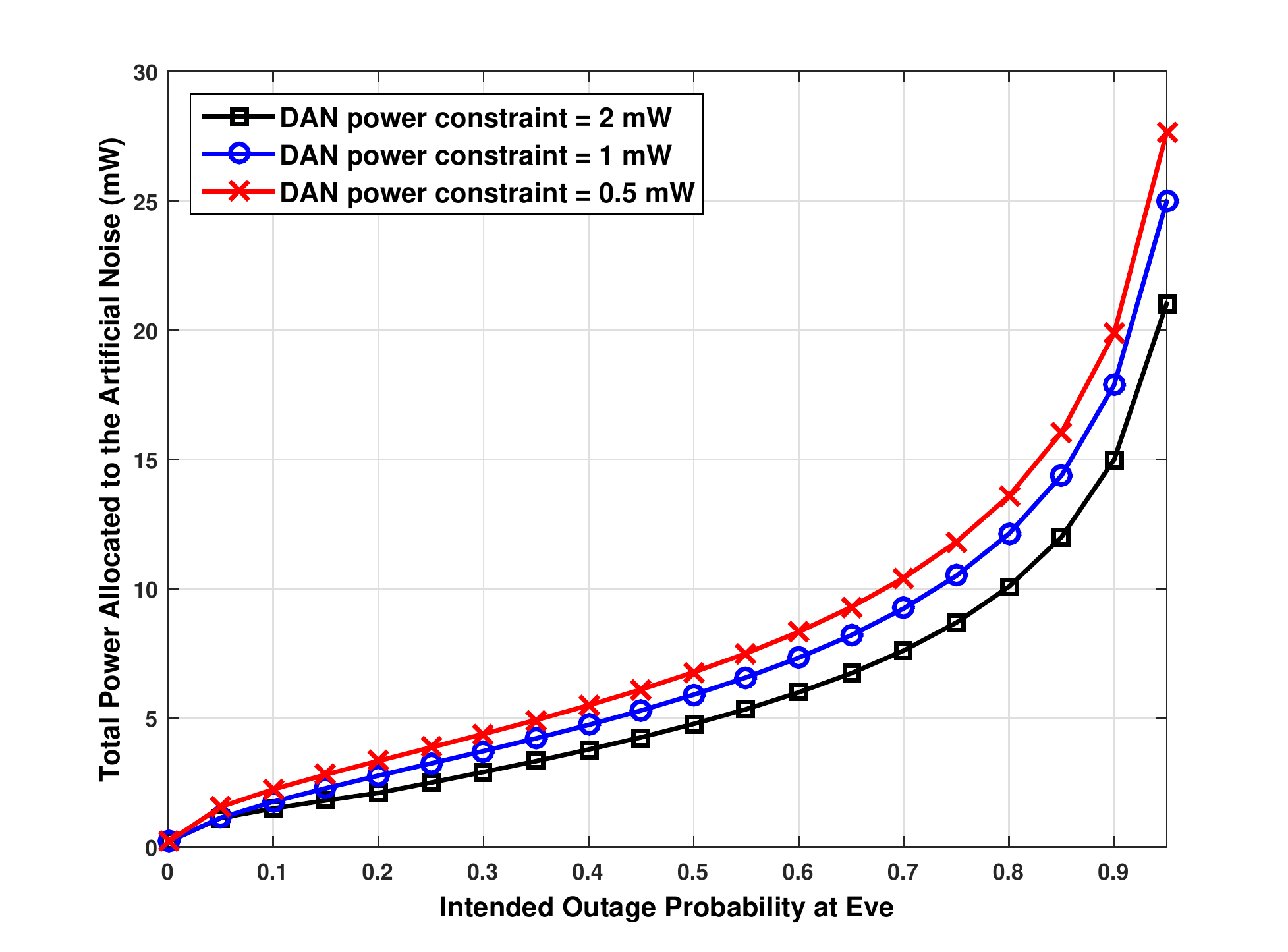}
\caption{The optimal power is needed for AN (total SAN and DAN) versus target outage probability ($\beta$) at Eve, while there is a power constraint at the Destination}
\label{fig:constraint}
\end{figure}

Although \emph{Full Duplex} technique is advancing so fast and today it is possible to cancel even high levels of \emph{Self-Interference} at the destination (\cite{Co-Channel:2010} and \cite{Breaking:2013}), but it is still useful to investigate how power consumption will be changed by limiting the maximum power allocated to the DAN. This is the case that is considered in Fig.~\ref{fig:constraint}. As we see in this figure, at the outage probability $\beta=0.9$ and for $N_{E}=N_{A}=N_{B}=4$, when DAN power limitation decreases from $2 mW$ to $0.5 mW$, total required Power for AN increases from $15 mW$ to $20 mW$.

\section{Conclusion}

In this paper, using destination artificial noise along with source artificial noise is introduced to guarantee an intended outage probability at Eve, while ensuring a certain SNR at Bob. For both constrained and unconstrained power allocation scenarios it has been shown how using the DAN, decreases the total power which is required to guarantee an intended outage probability at Eve. Using outage probability approach instead of secrecy capacity, makes our solution appropriate for the quasi-static channels.

\bibliographystyle{spbasic}
\bibliography{Paper-1_Refs_New}
\end{document}